\documentclass[letterpaper,10pt]{article}

\usepackage{amsmath}
\usepackage{amssymb}
\usepackage{graphicx}
\usepackage{bm}
\usepackage{epsfig}
\usepackage{amsfonts}
\usepackage{color}
\usepackage{multicol}

\newcommand{\beq}{\begin{equation}}
\newcommand{\eeq}{\end{equation}}
\newcommand{\bea}{\begin{eqnarray}}
\newcommand{\eea}{\end{eqnarray}}
\newcommand{\beas}{\begin{eqnarray*}}
\newcommand{\eeas}{\end{eqnarray*}}

\renewcommand{\vec}{\boldsymbol}
\newcommand{\be}{\begin{equation}}
\newcommand{\ee}{\end{equation}}
\newcommand{\bear}{\begin{eqnarray}}
\newcommand{\eear}{\end{eqnarray}}
\newcommand{\ba}{\begin{array}}
\newcommand{\ea}{\end{array}}

\begin{document}

\title{From the chiral magnetic wave to \\ the charge dependence of elliptic flow}

\author{Y. Burnier,$^1$ D. E. Kharzeev,$^{2,3}$ J. Liao,$^{4,5}$ H.-U. Yee$^2$
\vspace{3mm}
\\ 
\it{\normalsize $^1$Institute for Theoretical Physics, University of Bern} \\
\it{\normalsize Sidlerstrasse 5, CH-3012 Bern, Switzerland}\\
\it{\normalsize $^2$Department of Physics and Astronomy, Stony Brook University, } \\
\it{\normalsize Stony Brook, New York 11794-3800, USA}\\
\it{\normalsize $^3$Department of Physics, Brookhaven National Laboratory,}\\ 
\it{\normalsize Upton, New York 11973-5000, USA}\\
\it{\normalsize $^4$Physics Department and Center for Exploration of Energy and Matter,}\\
\it{\normalsize Indiana University, 2401 N Milo B. Sampson Lane,}\\
\it{\normalsize Bloomington, IN 47408, USA}\\
\it{\normalsize $^5$RIKEN BNL Research Center, Bldg. 510A,  } \\
\it{\normalsize Brookhaven National Laboratory, Upton, NY 11973, USA}
}




\date{\today}

\maketitle

\begin{abstract}
The quark-gluon plasma formed in heavy ion collisions contains charged chiral fermions evolving in an external magnetic field. 
At finite density of electric charge or baryon number (resulting either from nuclear stopping or from fluctuations), the triangle anomaly induces in the plasma the Chiral Magnetic Wave (CMW). The CMW first 
 induces a separation of the right and left chiral charges along the magnetic field; the resulting dipolar axial charge density in turn induces the oppositely directed vector charge currents leading to an electric quadrupole moment of the quark-gluon plasma. Boosted by the strong collective flow, the electric quadrupole moment translates into the charge dependence of the elliptic flow coefficients, so that $v_2(\pi^+) < v_2(\pi^-)$ (at positive net charge). Using the latest quantitative simulations of the produced magnetic field and solving the CMW equation, 
 we make further quantitative estimates of the produced $v_2$ splitting and its centrality dependence. We compare the results with the available experimental data.
\end{abstract}


\section{Introduction}
In (off-central) heavy ion collisions, the moving ions produce a huge magnetic field that could reveal an interesting dynamics in the plasma. The chiral anomaly present in quantum electrodynamics (QED) induces interesting ``macroscopic" effects in the plasma:  the magnetic field together with the presence of a non-vanishing electric or axial charge density lead to axial or electric currents. These currents lead respectively to the so-called Chiral Separation Effect (CSE) \cite{son:2004tq,Metlitski:2005pr,Gorbar:2011ya} and the Chiral Magnetic Effect (CME) \cite{Kharzeev:2004ey,Kharzeev:2007tn,Kharzeev:2007jp,Fukushima:2008xe,Kharzeev:2009fn}.

The CME needs an initial axial charge fluctuation. It can be created by sphaleron transitions that are essentially random, i.e. the sign of the effect varies from event to event. Note that the QCD sphaleron rate is expected to be large ($\Gamma\sim0.3\; T^4$) \cite{Moore:2010jd} and numerous sphaleron transitions happen in a single collision. The presence of strong magnetic field was found to increase the sphaleron rate \cite{Basar:2012gh}, but for the realistic 
values of magnetic field in heavy ion collisions the change in the rate is small. 
If we suppose that sphalerons create a non-zero axial chemical potential $\mu_A$, the QED anomaly induces a vector current $j_V^i = \bar{\psi} \gamma^i \psi$ in the presence of an external magnetic field (CME):
\be
\vec j_V=\frac{N_c\ e}{ 2\pi^2} \mu_A \vec B. \label{cme}
\ee

The vector current\footnote{The existence of this current in the quark-gluon plasma has been established in lattice QCDxQED \cite{Yamamoto:2011gk}.} induces a movement of electric charges leading to an electic dipole in the plasma i.e. to ``local parity violation".
Recently, STAR \cite{:2009uh, :2009txa}, PHENIX \cite{phenix} and ALICE \cite{Selyuzhenkov:2011xq} collaborations observed charge asymmetry fluctuations possibly providing an evidence for the CME. The interpretation of the results however is still under intense discussion, see e.g. \cite{BKL,Review,Kharzeev:2010gr}.
\begin{figure}
\begin{center}
\includegraphics[width=70mm, height=50mm]{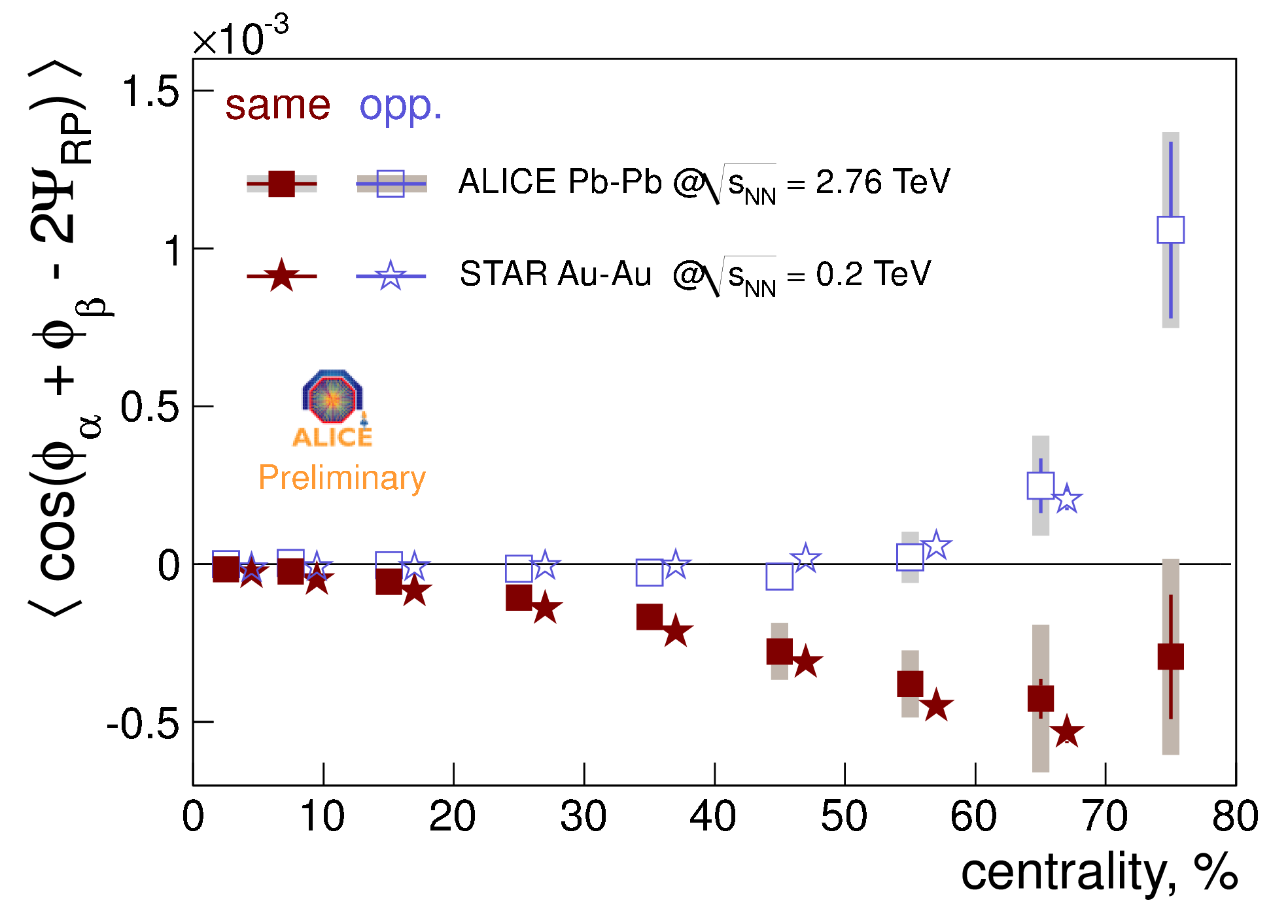}
\caption{Charged pair azimuthal correlations measured by the STAR at RHIC and the ALICE at LHC from Ref.~\cite{Selyuzhenkov:2011xq}.}
\end{center}
\end{figure}

The CSE requires an initial density of vector charges e.g.\ a finite baryon or electric charge number density,
which can be present in the plasma not just as a fluctuation but as a non-zero background.
  Together with a magnetic field, the QED anomaly leads to an axial current:
\be
\vec j_A=\frac{N_c\ e}{ 2\pi^2} \mu_V \vec B , \label{cse}
\ee
where $\mu_V$ is the vector chemical potential.

The coupling of the vector and axial currents induced by the anomaly in the presence of magnetic field brings to the existence a novel type of collective gapless excitation 
in plasma - the Chiral Magnetic Wave (CMW) \cite{Kharzeev:2010gd}. While the CMW can exist even at zero density of vector charge, it leads to a very 
interesting phenomenon when the finite vector charge density is present \cite{Burnier:2011bf}. One can visualize it in the following way: first the CSE 
 leads to a dipolar separation of axial charges along the $\vec B$ field direction; then the finite density of axial charge in turn induces the oppositely directed CME currents resulting  
in a charge quadrupole that has the same sign in all collisions (as long as the density of vector charge is positive). Consequently there are more positive charges at the poles of almond-shape fireball (since $\vec B$ is primarily out-of-plane) than at the equator (in the reaction plane). This eventually gives rise to the difference in elliptic flows between positive charged particles and the negatively charged ones.
In other words, the combination of CME and CSE, that is the chiral magnetic wave, leads to a charge dependence of the elliptic flow that survives even  after averaging over events. 

\section{The charge dependence of elliptic flow}

We will estimate here the size of the charge quadrupole created by the chiral magnetic wave in heavy ion collisions. 

\subsection{Chiral magnetic wave}
The co-evolution of the vector and axial currents can be described in the framework of Chiral Magnetic Wave (CMW) equation \cite{Kharzeev:2010gd}. Let us first give a short review of its derivation.
We can rewrite the anomaly formulas (\ref{cme},\ref{cse}) as
\beq
\left(\begin{array}{c} \vec j_V \\ \vec j_A\end{array}\right) = \frac{N_c\;e \vec B}{2\pi^2} \left(\begin{array}{cc}
0 & 1 \\ 1 & 0
\end{array}\right) \left(\begin{array}{c}
\mu_V \\ \mu_A
\end{array}\right). \label{cmw1}
\eeq
On the other hand, the chemical potential depends on the currents:
\bea
\left(\begin{array}{c}
 \mu_V \\ \mu_A
\end{array}\right)&=&\left(\begin{array}{cc}
\frac{\partial\mu_V}{ \partial j^0_V} & \frac{\partial\mu_V}{ \partial j^0_A} \\
 \frac{\partial\mu_A}{ \partial j^0_V} & \frac{\partial\mu_A}{\partial j^0_A }
\end{array}\right) \left(\begin{array}{c}
 j^0_V \\ j^0_A 
\end{array}\right) + {\cal O}\left(\left(j^0\right)^2\right) \\&\equiv&
\left(\begin{array}{cc}
\alpha_{VV}& \alpha_{VA}\\
 \alpha_{AV}  & \alpha_{AA} 
\end{array}\right) \left(\begin{array}{c}
 j^0_V \\ j^0_A 
\end{array}\right)+ {\cal O}\left(\left(j^0\right)^2\right)
.
\eea
Remembering that the chemical potentials are derivatives of the free energy:
$$\quad
\mu_i = \frac{\partial {\cal F}}{ \partial j^0_i} ,\quad i=V,A,
$$
we can define the susceptibility matrices $\alpha$:
\beq
\alpha_{ij} = \frac{\partial^2 {\cal F}}{ \partial j^0_i \partial j^0_j}.
\eeq
Under parity transformation $V\to - V$ and $A\to A$, so that 
$\alpha_{VA}=\alpha_{AV}=0$.
Moreover one can show that $\alpha_{VV}\sim \alpha_{AA}\equiv \alpha$ in the large $N_c$ limit. Inserting these results back into equation (\ref{cmw1}), we get
\beq
\left(\begin{array}{c} \vec j_V \\ \vec j_A\end{array}\right) = \frac{N_c\;e \vec B \alpha}{2\pi^2} \left(\begin{array}{cc}
0 & 1 \\ 1 & 0
\end{array}\right) \left(\begin{array}{c}
j^0_V \\ j^0_A
\end{array}\right).\label{semifinal}
\eeq
We can rewrite these equations in the basis of left and right spinors:
\beq
\vec j_{L,R}= \mp \left(\frac{N_c\;e \vec B \alpha}{2\pi^2}\right) j^0_{L,R} +\cdots \label{consti2}
\eeq
The final step is to combine the equation (\ref{consti2}) with the conservation laws $\partial_\mu j^\mu_{L,R}=0$ that are approximately valid in time scales much shorter than the quark mass; this is easily satisfied in real experiments. On general grounds one should also include diffusion, which arises at the next order in the gradient expansion.
Denoting $x_1$ the direction of the magnetic field, the evolution of the CMW is described by the wave equation \cite{Kharzeev:2010gd}
\beq
\left(\partial_0 \mp{v} \partial_1 -D_L \partial^2_1 -D_T\partial^2_T\right) j^0_{L,R}=0,
\eeq
with $v=\frac{N_c e B \alpha}{2\pi^2}$ the velocity of the wave and $D_L$ ($D_T$) the longitudinal (transverse) diffusion constant. 
Note that the constants $\alpha,~D_L,~D_T$ are $B$ and $T$ dependent and such that $v<c$.
We use in the following the values of $v_\chi^f$ and $D_L^f$ that were computed in ref.~\cite{Kharzeev:2010gd} in the framework of the Sakai-Sugimoto model in the large $N_c$ quenched approximation. 

To describe the initial "almond" configuration of the QCD matter we use the KLN model \cite{Kharzeev:2001gp} based on parton saturation and $k_T$ factorization. $Au-Au$ collisions have been simulated, with realistic Woods-Saxon nuclear densities.
The axial chemical potentials at initial time are set to zero.

\begin{figure}
\begin{center}
\includegraphics[width=50mm, height=50mm]{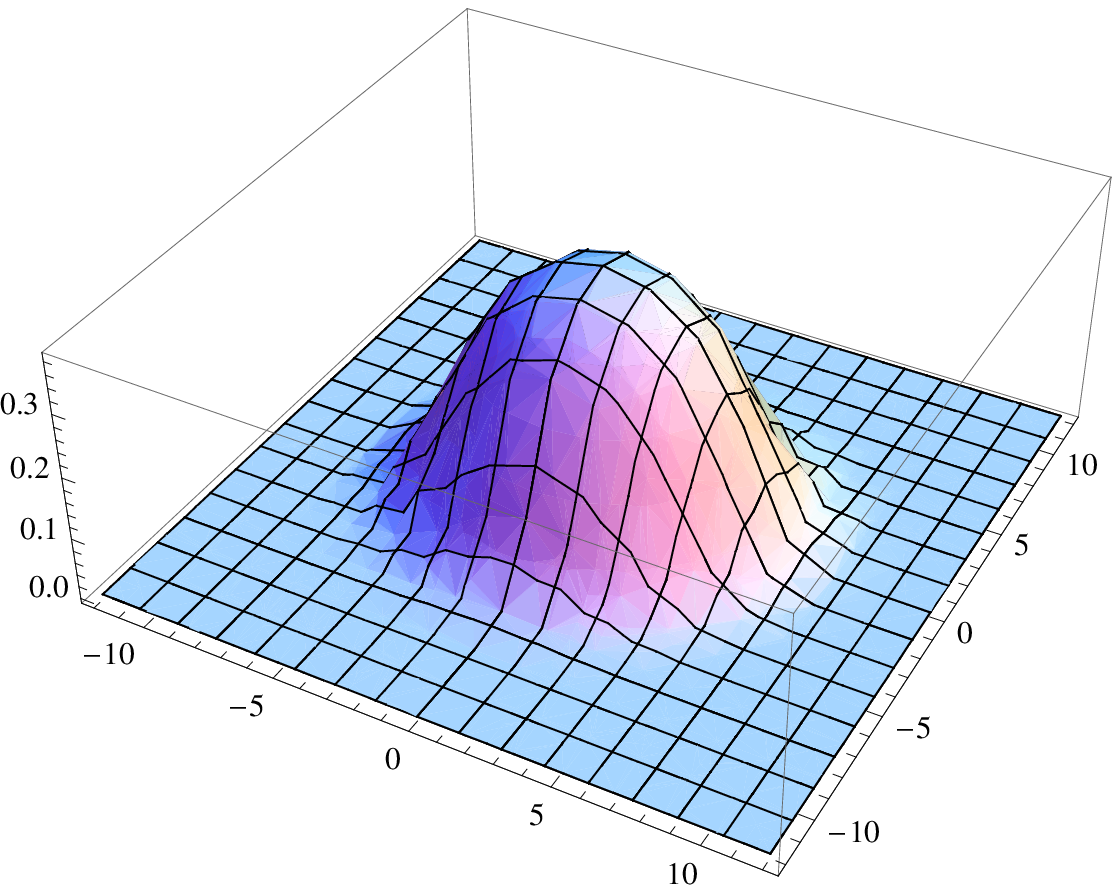}
\includegraphics[width=60mm, height=60mm]{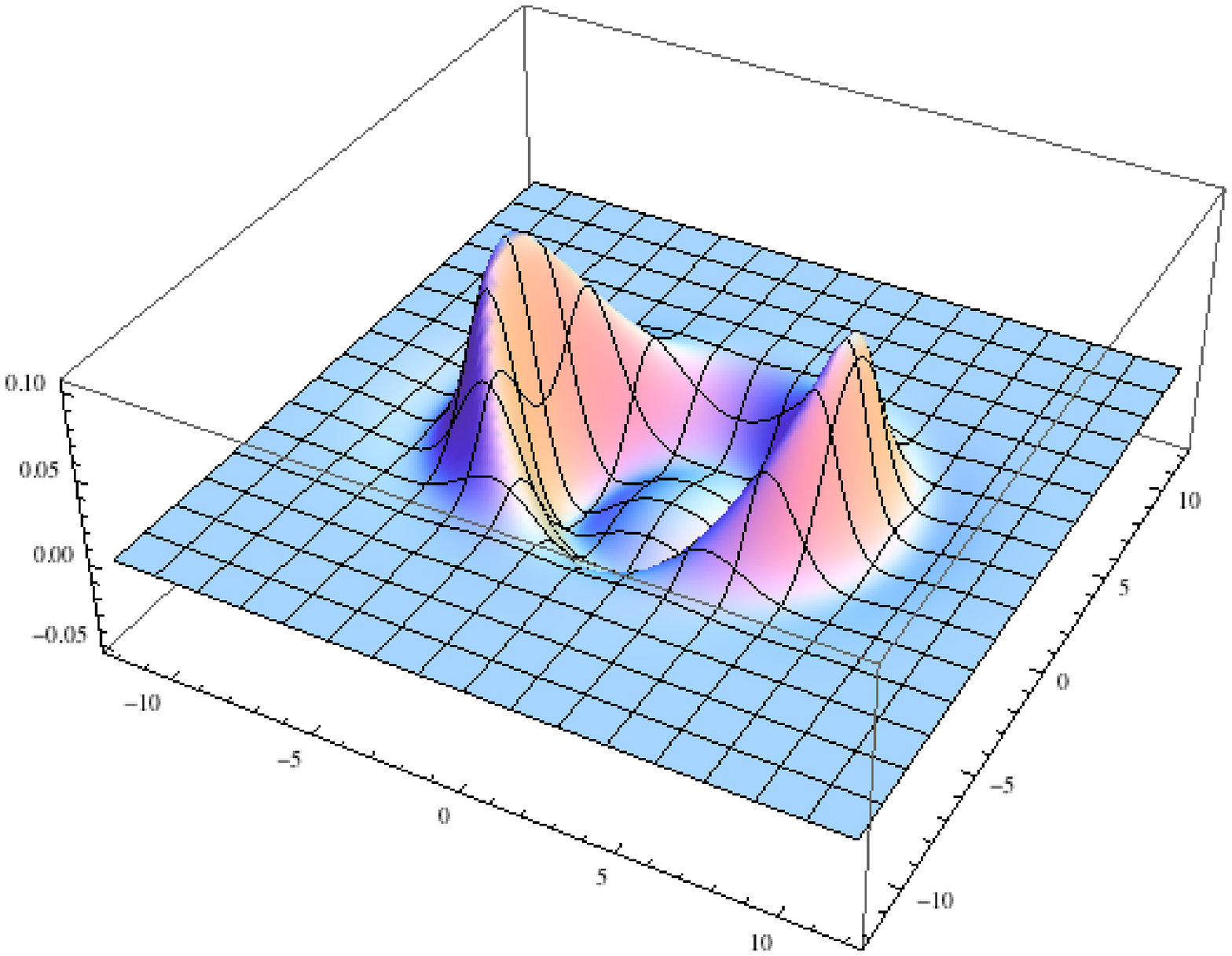}\label{fig:CMW}
\caption{Initial baryon density and evolution of the excess of charges in the plane transverse to the beam axis. For clarity, we have subtracted the charge density distribution without the CMW in the second picture. Magnetic field strength used $eB=2.8 m_\pi^2$, lifetime of magnetic field $\tau=4$ fm,  temperature $T=165$ MeV, impact parameter $b=3$ fm.}
\end{center}
\end{figure}

We consider the propagation of the $u$ and $d$ quark currents in terms of two different CMWs. We omit here the strange quark as it has no net density in the plasma. As the quark flavours have different charges $q_f$, we use $v(q_f B), D_{L,T}(q_f B)$ as coefficients for the CMW. The total electric charge distribution is obtained by summing the different waves as
\be
j^0_e=\sum_f q_f \left(j^{0,f}_{L}+j^{0,f}_R\right).
\ee
The resulting distribution is shown in fig. \ref{fig:CMW}.  We see that the excess of positive charges move towards the pole of the fireball whereas the center contains less positive charges, leading to a charge quadrupole.

\subsection{From charge quadrupole to  elliptic flow}

In principle the CMW evolves in an expanding background. A full computation would require rewriting the CMW in an expanding medium and performing a full hydrodynamical simulation. 
Here for a first estimate of the effect, we will make the following approximations. We first compute the charge distribution of the wave evolution in a fixed background, and then include the effect of the flow. This would give precise results if the magnetic field has a short lifetime. In the case of a long lifetime, the propatation of the wave and the expansion of the medium  will  get entangled, which is a more complicated case that we will further study elsewhere. The dominant flow effect is the strong radial flow that correlates the emitted particle's transverse momentum magnitude and direction with its spatial radial position and azimuthal angle. We thus assume that the charge asymmetry distribution (at the end of CMW evolution) is simply carried over by the radial flow.

\begin{figure}
\begin{center}
\includegraphics[width=45mm, height=40mm]{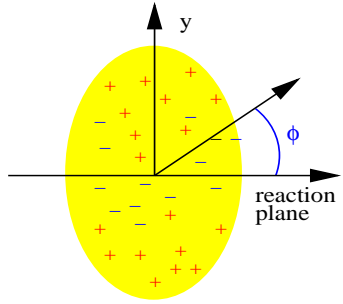}\label{fig:3}
\caption{Schematic demonstration of charge quadrupole boosted by strong collective (radial) flow.}
\end{center}
\end{figure}

After the evolution of the CMW, the azimuthal distribution of net charge density clearly  bears quadrupole component, i.e. the quark-gluon plasma becomes an electric quadrupole, see Fig.~\ref{fig:3}. 
As mentioned above, the strong radial flow aligns the particles' momenta along the direction of the flow. 
Therefore the net charge detected at an angle $\phi$ is proportional to 
$$\frac{d(N_+-N_-)}{d\phi} \propto \int r\, dr\, j^0_e(r,\phi).$$
The azimuthal anisotropy of net charge distribution is dominated by the elliptic component (i.e. the 2nd harmonics), and we approximately have
\beq
\frac{d(N_+-N_-)}{d\phi}=(\bar N_+-\bar N_-)[1-r_e\cos(2\phi)],\label{dndp}
\eeq
with $\bar N_\pm$ the total number of $\pm$ charged particles
and $r_e$ the ratio between the 2nd harmonic of the net charge   distribution to the total net charge:
\bea 
&&r_e=2\frac{q_e}{\rho_e},  \\ 
&&q_e=-\int r\, dr\, d\phi\, \cos (2\phi)\left(j_e^0(r,\phi)-j^0_{e,B=0}(r,\phi)\right),  \notag \\
&& \rho_e=  \int r \, dr \, d\phi\, j^0_e(r,\phi). \nonumber
\eea
Note that in the definition of $q_e$ we focus on the charge quadrupole entirely due to the CMW evolution by subtracting out the net charge distribution without CMW. Supposing that the hydrodynamical expansion of the plasma is not perturbed by the net charge distribution (which indeed is a very small perturbation to the overall bulk evolution), the total (charge independent) particle distribution will still follow the usual elliptic flow pattern  $[1+2v_2\cos(2\phi)]$. Using Eq.(\ref{dndp})  we can then write down the positive/negative charge distributions that include both the usual elliptic flow effect and the charge quadrupole contribution (up to   $\mathcal{O}(v_2 \times r_e\, A)$):
\beq
\frac{dN_\pm}{d\phi}=\bar N_\pm\left[1+(2v_2\mp r_e\, A ) \cos(2\phi)\right],\notag
\eeq
where 
$A=\frac{\bar N_+-\bar N_-}{\bar N_+ +\bar N_-}$ is the net charge asymmetry. It is then obvious to see that the elliptic flow is charge-dependent  $v_2^{\pm}= v_2 \mp r_e\, A/2$ with the 
splitting being
\beq
v_2^--v_2^+= r_e\,A \,.
\eeq 
The splitting is linear in the net charge asymmetry $A$, with the slope $r_e$ determined from the net charge distribution due to the CMW evolution.

\subsection{Evolution of the charges after the plasma phase}

Upon hadronization, the electric charges carried by the quarks are transferred into pions, kaons, protons and other hadrons. However, various hadronic reactions in the "afterburner" phase of the heavy ion collision may considerably alter the charge asymmetry pattern from the plasma phase. Protons and antiprotons as well as kaons and antikaons have quite different cross sections in the hadronic matter at finite baryon density, and this may well mask the effect one would like to look for.  Positive and negative pions seem to be the best channel, as they have a relatively small difference in the absorption cross sections and are produced in large quantities.  Therefore we suggest to measure the electric quadrupole moment of the plasma by using the difference of the elliptic flows of pions, $v_2(\pi^-) - v_2(\pi^+)$ as the observable.

\section{Results and comparison with experiments}

As already discussed, the creation of the charge quadrupole from CMW requires an initial vector charge density. In relatively low-energy heavy ion collisions the plasma naturally acquires sizable initial baryonic charge density (and a relatively large charge asymmetry $A$) due to nuclear stopping effects. Therefore 
it is natural to first look for the predicted  elliptic flow difference  $v_2(\pi^-) - v_2(\pi^+)=r_e A$ in low energy collisions, as proposed and estimated in \cite{Burnier:2011bf}. At high energy, $A$ becomes small on average as the initial baryon density is small. However, as a result of event-by-event fluctuations, the charge asymmetry $A$  of hadrons measured in a limited kinematical domain (e.g. in a slice of rapidity) always fluctuates and can be quite sizable in a fraction of events. Therefore a second way to look for the predicted $v_2$ splitting is to  measure the  $v_2(\pi^-) - v_2(\pi^+)$ versus the charge asymmetry $A$ in high energy collision events (by binning them according to the charge asymmetry). In that way one can (1) test the predicted linear relation between the splitting and the charge asymmetry, and (2) extract the slope that can be directly compared with the quadrupole quantity $r_e$ from CMW computation. This very useful strategy has been recently developed and used by the STAR collaboration to measure the charge quadrupole at high collision energies \cite{STAR_v2}.

\subsection{Results from the numerical simulation}

\begin{figure}
\begin{center}
\includegraphics[width=70mm]{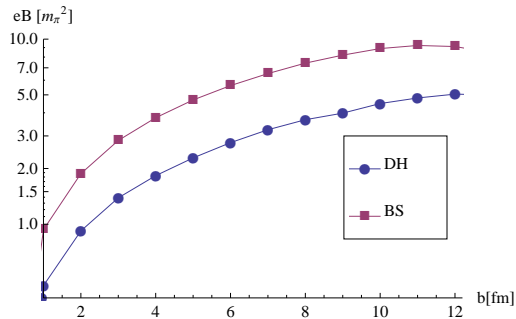}\label{figeB}
\caption{Magnetic field from recent event-by-event computations in Ref. \cite{Bzdak:2011yy} (BS) and Ref. \cite{Deng:2012pc} (DH).}
\end{center}
\end{figure}

In the original estimate \cite{Burnier:2011bf} of the quadrupole moment, we assumed that 
the initial up and down contents of the plasma followed the up and down quark content of the gold ion. This was motivated by the picture of baryon stopping appropriate at low energies. Here we present slightly different results, where we have fixed the initial charge density and realize it with equal amount of up and down quarks. This is motivated by  comparison with the STAR results at high energies where the amount of charge was measured directly. The resulting difference is however very small numerically  and is well below other sources of incertitude.
The main difference with the simulation of Ref.\cite{Burnier:2011bf} is that we have now used the state-of-art computation (on event-by-event basis) of the magnetic field (shown in Fig.\ref{figeB}) from Bzdak and Skokov (BS) \cite{Bzdak:2011yy}, as well as from Deng and Huang (DH) \cite{Deng:2012pc} instead of the analytical estimate of magnetic field from Ref.\cite{Kharzeev:2007jp}. 
Note that due to differences in their implementation, the results of BS are larger by a factor 2 than the results of DH, but both are significantly larger than the estimate of Ref.\cite{Kharzeev:2007jp} for large impact parameters. 

\begin{figure}
\begin{center}
\includegraphics[width=100mm]{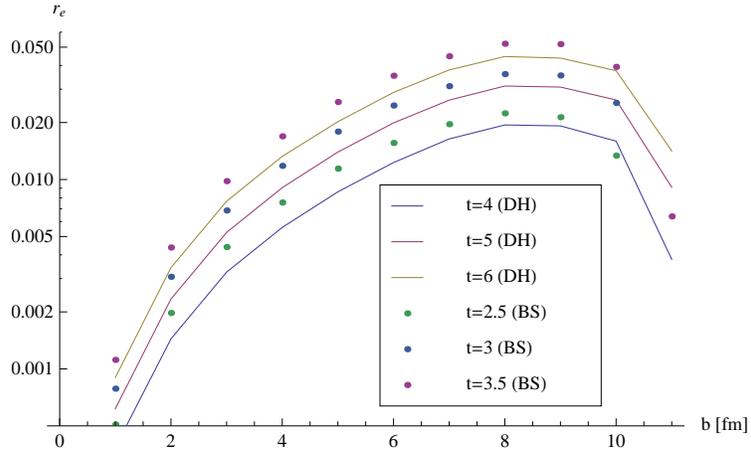}\label{fig56}
\caption{The  magnitude of the charge quadrupole $r_e$ computed using the event-by-event evaluated magnetic field with a lifetime $t$ [fm/c].}
\end{center}
\end{figure}

The time evolution of the magnetic field in the plasma critically depends on the medium's transport properties. If the electric conductivity is large, magnetohydrodynamics is applicable and the magnetic flux is conserved. In the opposite limit, the plasma effects are negligible and the magnetic field decreases quickly as the spectator ions fly away. An estimate of the effect of electric conductivity on the lifetime of the plasma has been presented in \cite{Tuchin:2010vs}; 
a more detailed study of this issue is clearly needed. Since the overall effect generated by the magnetic field would be integrated over time, instead of convoluting with the complicated time-dependent $\vec B$ field,  we will  compute the magnitude of the charge quadrupole by assuming the magnetic field  to stay constant from $t_0=0$ till the end of the $\vec B$ field lifetime. The resulting electric quadrupole moment $r_e$ is shown in Fig.~\ref{fig56} as function of the impact parameter $b$. 

\subsection{Experimental results}

The first experimental evidences for the (global) electric quadrupole effect were found for 0-80\% centrality events by the STAR collaboration at RHIC \cite{Mohanty:2011nm} and the CERES collaboration at SPS \cite{Adamova:2012md}.
\begin{figure}
\begin{center}
\includegraphics[width=65mm, height=40mm]{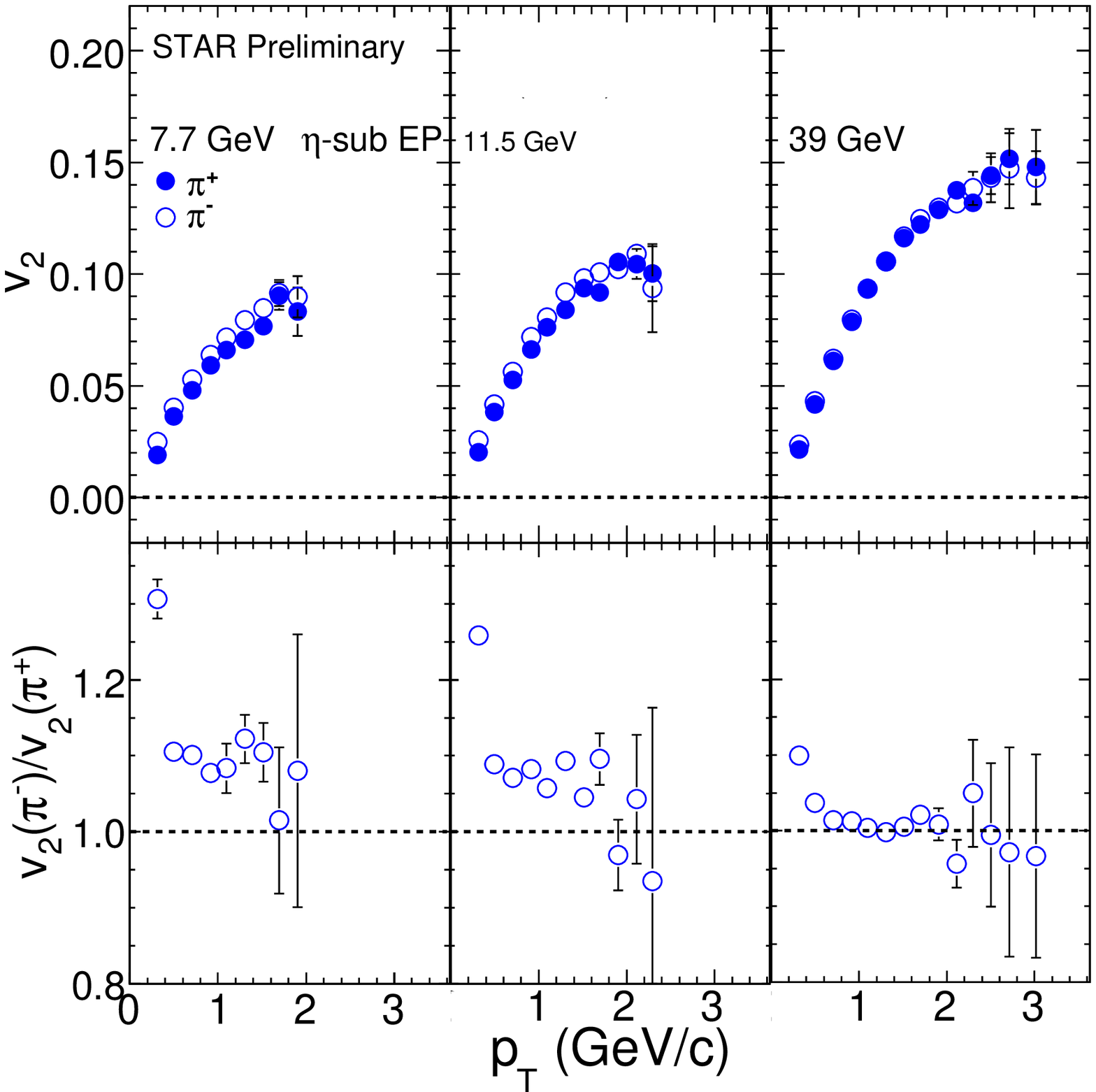}
\includegraphics[width=35mm, height=40mm]{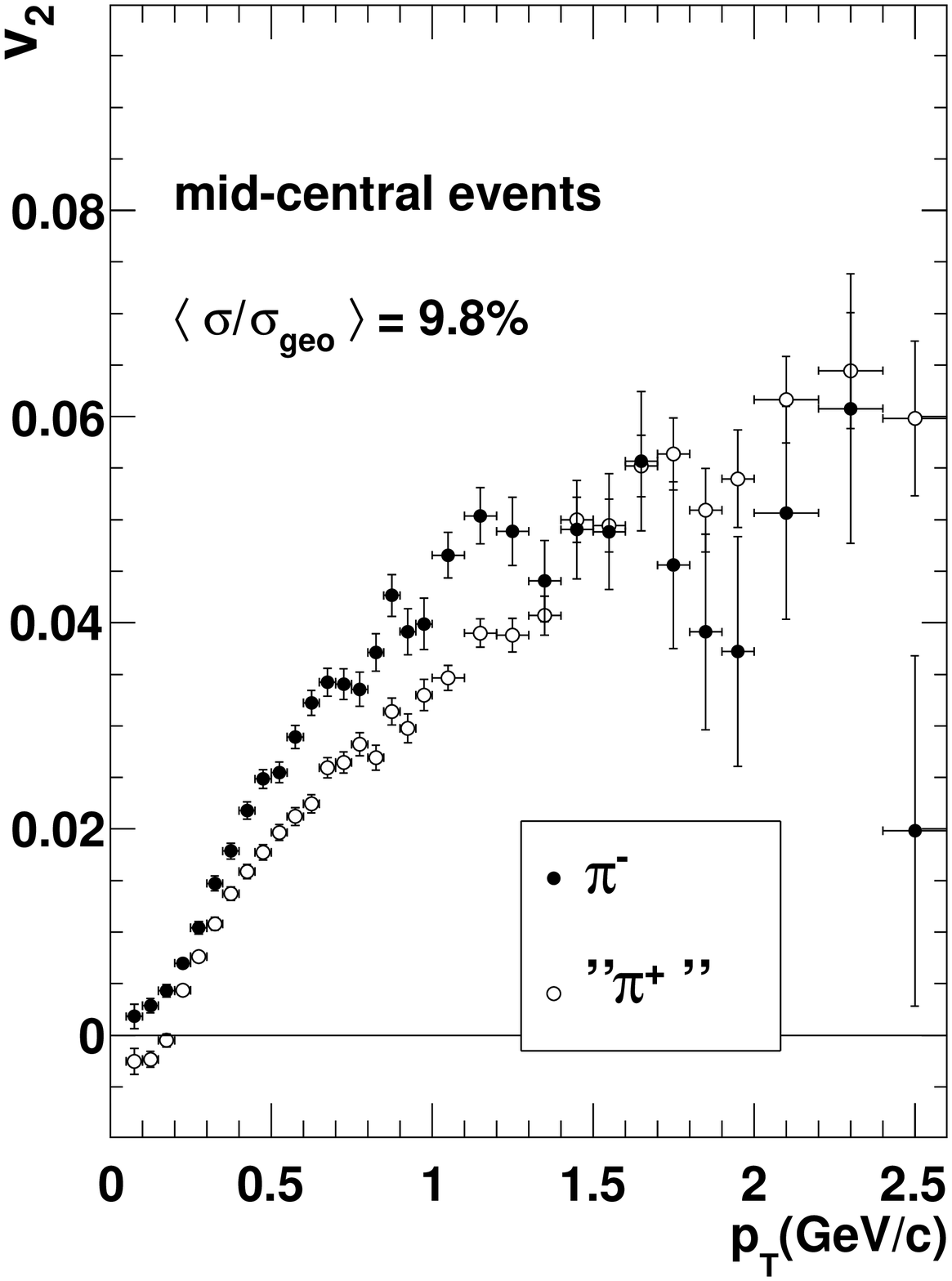}
\caption{The charge dependence of the elliptic flow from minimum bias measurements by the STAR Beam Energy Scan at RHIC  and by the CERES at SPS.}
\end{center}
\end{figure}
The sign and the magnitude of the difference in the positive/negative pion $v_2$ appear to agree with the prediction \cite{Burnier:2011bf}. For instance for $\sqrt{s}=11 \, \rm GeV$ collisions at RHIC, with $A\sim 0.3$ and $r_e\sim 0.01$ for $t=4$ fm using the magnetic field of (DH), one expects the difference in elliptic flow $\sim r_e A\sim 0.003$ (with $v_2$ of $\pi^-$ bigger than $\pi^+$), which is 10\% of the overall elliptic flow at this energy. These observations provide an important evidence for the proposed CMW effect. Note however that at such low collision energies, the precise interpretation of data might get difficult due to a short plasma phase lifetime, a significant contribution from hadronic phase, as well as due to a possible effects of stopping on the bulk evolution \cite{Steinheimer:2012bn}.

Very recently the STAR collaboration completed an even more interesting analysis at high energy heavy ion collisions (up to $\sqrt{s}=200\,\rm GeV$) following the second method mentioned above \cite{STAR_v2}. By measuring the difference in elliptic flow of positive/negative pions for events binned according to their charge asymmetry $A$ (at a given centrality class), they found a linear dependence of the splitting $v_2^- -v _2^+= r_e A$ on the charge asymmetry $A$.  They extracted the slope of this linear dependence, that is to be compared with the magnitude of the charge quadrupole $r_e$. 
The comparison between the STAR measured slope parameter and our computed charge quadrupole $r_e$ shows very good agreement over a wide range of centrality for the magnetic field with DH magnitude and  lifetime of $t \simeq 5$ fm/c or with BS magnitude and lifetime between $t=2.5-3$ fm/c.  These measurements at high collision energies and the success of our prediction provide a strong evidence for the proposed electric charge quadrupole induced by the chiral magnetic wave.

\section{Outlook}

To summarize, we have shown that the chiral magnetic wave induces an electric quadrupole deformation of the quark-gluon plasma created in heavy ion collisions, and leads to the observable splitting of elliptic flows of positive and negative pions. Quantitative estimates of the magnitude and sign of the effect (at both low and high collision energies) as well as its linear dependence on the charge asymmetry and its centrality dependence (at high collision energies) seem to match the experimental data.  These results in our opinion can be viewed as a strong evidence for the CMW-induced quadrupole deformation of the quark-gluon plasma. 

While these developments are very encouraging, more detailed studies still have to be done.
On the theoretical side, a more sophisticated computation of the CMW needs to  be performed in an expanding background and taking into account the spatial and temporal dependences of the temperature and magnetic field. The modeling of the medium and its  expansion dynamics can be significantly improved by coupling the CMW evolution with the well-developed hydrodynamic simulations. We emphasize that   
the magnetic field and its lifetime are still the main incertitudes that bear significant influence on the quantitative results. Progress was made recently on the determination of the electric conductivity of a pure gluon plasma \cite{Ding:2010ga,Burnier:2012ts} and we hope that these calculations will be extended to the real quark-gluon plasma. Once the electric conductivity is known, it would be interesting to perform dedicated simulations of the magnetic field induced inside the plasma. On the experimental side, measurements can be developed to further test the proposed $v_2$ splitting. For example, from the recently completed U+U collisions, one can look for events with very small impact parameter but sizable elliptic flow (the so-called ``body-body'' collisions) \cite{Voloshin:2010ut}. These events are believed to have negligible magnetic field and therefore no effect from CMW-induced charge quadrupole. Therefore if these events could be selected successfully, then  one can test the disappearance of the $v_2$ splitting. One may also test the linear relation of $v_2$ splitting on the charge asymmetry $A$ using the data from the Beam Energy Scan. It would also be extremely interesting to check if this effect, with a linear relation between $v_2$ and $A$, still persists at the LHC energy. 

\section*{Acknowledgments}
We are grateful to  Wei-Tian Deng and Xu-Guang Huang, and Adam Bzdak and Vladimir Skokov for discussions and for providing the results of their computation of the magnetic field. 
We also thank Aihong Tang, Gang Wang and Hongwei Ke for useful discussions and communications. 




\begin{thebibliography}{99} \frenchspacing


\bibitem{son:2004tq}
  D.~T.~Son and A.~R.~Zhitnitsky,
  Phys.\ Rev.\  D {\bf 70}, 074018 (2004).

\bibitem{Metlitski:2005pr}
  M.~A.~Metlitski and A.~R.~Zhitnitsky,
  Phys.\ Rev.\  D {\bf 72}, 045011 (2005).

\bibitem{Gorbar:2011ya}
  E.~V.~Gorbar, V.~A.~Miransky and I.~A.~Shovkovy,
  Phys.\ Rev.\ D {\bf 83} (2011) 085003.


\bibitem{Kharzeev:2004ey}
  D.~Kharzeev,
  Phys.\ Lett.\  B {\bf 633}, 260 (2006).

\bibitem{Kharzeev:2007tn}
  D.~Kharzeev and A.~Zhitnitsky,
  Nucl.\ Phys.\  A {\bf 797}, 67 (2007).


\bibitem{Kharzeev:2007jp}
  D.~E.~Kharzeev, L.~D.~McLerran and H~.J.~Warringa,
  Nucl. Phys.  A {\bf 803}, 227 (2008).

\bibitem{Fukushima:2008xe}
  K.~Fukushima, D.~E.~Kharzeev and H.~J.~Warringa,
  Phys.\ Rev.\  D {\bf 78}, 074033 (2008).

  \bibitem{Kharzeev:2009fn}
  D.~E.~Kharzeev,
  Annals Phys.\  {\bf 325}, 205 (2010).

\bibitem{Moore:2010jd}
  G.~D.~Moore and M.~Tassler,
  JHEP {\bf 1102} (2011) 105.

\bibitem{Basar:2012gh} 
  G.~Basar and D.~E.~Kharzeev,
  Phys.\ Rev.\ D {\bf 85}, 086012 (2012)
  [arXiv:1202.2161 [hep-th]].

\bibitem{Yamamoto:2011gk} 
  A.~Yamamoto,
  Phys.\ Rev.\ Lett.\  {\bf 107}, 031601 (2011)
  [arXiv:1105.0385 [hep-lat]];
  Phys.\ Rev.\ D {\bf 84}, 114504 (2011)
  [arXiv:1111.4681 [hep-lat]].
  
\bibitem{:2009uh}
  B.~I.~Abelev {\it et al.}  [STAR Collaboration],
  Phys.\ Rev.\ Lett.\  {\bf 103}, 251601 (2009).

\bibitem{:2009txa}
  B.~I.~Abelev {\it et al.}  [STAR Collaboration],
  Phys.\ Rev.\ C {\bf 81}, 054908 (2010).

\bibitem{phenix}
N.~N.~ Ajitanand, S. Esumi, R.~A.~ Lacey [PHENIX Collaboration], in: Proc. of the RBRC Workshops, vol. 96, 2010.

\bibitem{Selyuzhenkov:2011xq}
  I.~Selyuzhenkov [ALICE Collaboration],
  Prog.\ Theor.\ Phys.\ Suppl.\  {\bf 193} (2012) 153.


\bibitem{BKL}
  A.~Bzdak, V.~Koch and J.~Liao,
  Phys.\ Rev.\  C {\bf 81}, 031901 (2010);
  Phys.\ Rev.\ C {\bf 83}, 014905 (2011).
J.~Liao, V.~Koch and A.~Bzdak,  Phys.\ Rev.\ C {\bf 82}, 054902.


\bibitem{Review} 
  A.~Bzdak, V.~Koch and J.~Liao,
  arXiv:1207.7327 [nucl-th].

\bibitem{Kharzeev:2010gr}
  D.~E.~Kharzeev, D.~T.~Son,
  Phys.\ Rev.\ Lett.\  {\bf 106}, 062301 (2011).



\bibitem{Kharzeev:2010gd}
  D.~E.~Kharzeev and H.~-U.~Yee,
  Phys.\ Rev.\ D {\bf 83} (2011) 085007.

\bibitem{Burnier:2011bf} 
  Y.~Burnier, D.~E.~Kharzeev, J.~Liao and H.~-U.~Yee,
  Phys.\ Rev.\ Lett.\  {\bf 107}, 052303 (2011).


\bibitem{Kharzeev:2001gp}
  D.~Kharzeev, M.~Nardi,
  Phys.\ Lett.\  {\bf B507}, 121-128 (2001);
  D.~Kharzeev and E.~Levin,
  Phys.\ Lett.\  B {\bf 523} (2001) 79
; 
  D.~Kharzeev, E.~Levin, M.~Nardi,
  Phys.\ Rev.\  {\bf C71}, 054903 (2005).

\bibitem{STAR_v2}
G.~Wang, talk at the ``RBRC Workshop on P- and CP-Odd Effects in Hot and Dense Matter'', BNL, June 2012. http://www.bnl.gov/pcp2012/

\bibitem{Bzdak:2011yy}
  A.~Bzdak and V.~Skokov,
  Phys.\ Lett.\ B {\bf 710} (2012) 171.
\bibitem{Deng:2012pc}
  W.~-T.~Deng and X.~-G.~Huang,
  Phys.\ Rev.\ C {\bf 85} (2012) 044907.
  
  
  
\bibitem{Mohanty:2011nm}
  B.~Mohanty [STAR Collaboration],
  J.\ Phys.\ G G {\bf 38} (2011) 124023.

\bibitem{Adamova:2012md}
  D.~Adamova {\it et al.}  [CERES Collaboration],
  arXiv:1205.3692 [nucl-ex].







\bibitem{Steinheimer:2012bn} 
  J.~Steinheimer, V.~Koch and M.~Bleicher,
  arXiv:1207.2791 [nucl-th].



  
\bibitem{Tuchin:2010vs} 
  K.~Tuchin,
  Phys.\ Rev.\ C {\bf 82}, 034904 (2010)
  [Erratum-ibid.\ C {\bf 83}, 039903 (2011)].
  
\bibitem{Ding:2010ga}
  H.~-T.~Ding, A.~Francis, O.~Kaczmarek, F.~Karsch, E.~Laermann and W.~Soeldner,
  Phys.\ Rev.\ D {\bf 83} (2011) 034504.
\bibitem{Burnier:2012ts}
  Y.~Burnier and M.~Laine,
  Eur.\ Phys.\ J.\ C {\bf 72} (2012) 1902.

\bibitem{Voloshin:2010ut} 
  S.~A.~Voloshin,
  Phys.\ Rev.\ Lett.\  {\bf 105}, 172301 (2010).





\end{thebibliography}
\end{document}